\documentclass[conference, 10pt]{IEEEtran}
\usepackage{hyperref}

\IEEEoverridecommandlockouts

\usepackage{fancyhdr}
\fancypagestyle{firstpageheader}{
    \fancyhf{}
    
    \fancyhead[C]{\Large \textbf{Extended Version of the Accepted Short Paper at The $43^{\mathrm{rd}}$ IEEE International Conference on Computer Design (ICCD) 2025, Dallas, TX, USA.}}
}

\usepackage{amsmath,amssymb,amsfonts}
\usepackage{graphicx}
\usepackage{textcomp}
\usepackage{diagbox}
\usepackage{booktabs}
\usepackage{colortbl}
\usepackage{multirow}
\usepackage[numbers,sort&compress]{natbib}
\usepackage{tabularx}
\usepackage{tabularray}
\usepackage{siunitx}
\usepackage{soul}
\usepackage{wrapfig}
\usepackage[caption=false]{subfig}
\usepackage[font=footnotesize,labelfont=bf]{caption}
\usepackage[normalem]{ulem}
\usepackage{algorithm}
\usepackage[noend]{algpseudocode}
\algtext*{EndWhile}
\algtext*{EndIf}
\algtext*{EndFor}
\algnewcommand\algorithmicforeach{\textbf{for each}}
\algdef{S}[FOR]{ForEach}[1]{\algorithmicforeach\ #1\ \algorithmicdo}
\usepackage{comment}
\usepackage{makecell}
\usepackage{xcolor}
\usepackage{enumitem}
\usepackage{footnote}  
\usepackage{pifont}

\definecolor{myblue}{HTML}{93CDDD}
\definecolor{mygreen}{HTML}{17891B}
\definecolor{myyellow}{HTML}{DCC60F}
\definecolor{myorange}{HTML}{EB801B}

\title{ECOLogic: \underline{E}nabling \underline{C}ircular, \underline{O}bfuscated, and Adaptive \underline{Logic} via eFPGA-Augmented SoCs}

\author{
    \IEEEauthorblockN{
        Ishraq Tashdid\IEEEauthorrefmark{1},
        Dewan Saiham\IEEEauthorrefmark{2},
        Nafisa Anjum\IEEEauthorrefmark{3},
        Tasnuva Farheen\IEEEauthorrefmark{3},
        Sazadur Rahman\IEEEauthorrefmark{1}
    }
    \IEEEauthorblockA{
        \IEEEauthorrefmark{1}Department of Electrical and Computer Engineering, University of Central Florida, USA \\
        \IEEEauthorrefmark{2}College of Optics and Photonics (CREOL), University of Central Florida, USA \\
        \IEEEauthorrefmark{3}Department of Computer Science and Engineering, Louisiana State University, USA \\
        ishraq.tashdid@ucf.edu, dewan.saiham@ucf.edu, nanjum2@lsu.edu, tfarheen@lsu.edu, mohammad.rahman@ucf.edu
    }
}

\begin{document}
\maketitle
\thispagestyle{firstpageheader} % Header only on first page

\begin{abstract}
%The accelerating demand for AI and machine learning hardware is driving a surge in application specific integrated circuit (ASIC) design, implementation, and fabrication, exacerbating carbon emissions, supply chain vulnerabilities, and hardware obsolescence. 
Traditional hardware platforms—ASICs and FPGAs—offer competing trade-offs among performance, flexibility, and sustainability. ASICs provide high efficiency but are inflexible post-fabrication, require costly re-spins for updates, and expose IPs to piracy risks. FPGAs offer reconfigurability and reuse, yet suffer from substantial area, power, and performance overheads, resulting in higher carbon footprints. We present \emph{ECOLogic}, a hybrid design paradigm that embeds lightweight eFPGA fabric within ASICs to enable secure, updatable, and resource-aware computation. Central to this architecture is \emph{ECOScore}, a quantitative scoring framework that evaluates IPs based on adaptability, piracy threat, performance tolerance, and resource fit to guide RTL partitioning. Evaluated across six diverse SoC modules, \emph{ECOLogic} retains an average of $90\%$ ASIC-level performance (up to $2$\,GHz), achieves $9.8$\,ns timing slack (versus $5.1$\,ns in FPGA), and reduces power by $480\times$ on average. Moreover, sustainability analysis shows a $99.7\%$ reduction in deployment carbon footprint and $300$--$500\times$ lower emissions relative to FPGA-only implementations. These results position \emph{ECOLogic} as a high-performance, secure, and environmentally sustainable solution for next-generation reconfigurable systems.

\end{abstract}
\vspace{-6pt}
\section{Introduction} 
\noindent Computing systems from microchips to data centers already emit $2.1–3.9\%$ of global greenhouse gases, a share that could approach $8\%$ within the next decade as digital demand soars \cite{henderson2020towards, freitag2021real, andrae2015global}.
Energy-efficient application-specific integrated circuits (ASICs) deliver high performance but cannot be updated after fabrication, requiring costly re-spins for any hardware modification, that increase carbon emissions and e-waste \cite{bartos2005asics, gupta2022act, tashdid2025safe}.
Field-programmable gate arrays (FPGAs) avoid re-fabrication through post-deployment reconfigurability, yet their large area, power, and performance overheads limit use in resource-constrained settings \cite{bartos2005asics}.
Modern ASIC supply chains also depend on untrusted foundries, exposing designs to intellectual-property theft, reverse engineering (RE), and counterfeiting \cite{guo2023evolute}. Meanwhile, ASICs lack self-repair, and FPGAs rarely meet stringent reliability or radiation-hardening requirements \cite{xu2021reliability}.
Together, these gaps call for a unified hardware architecture that offers ASIC-class efficiency, FPGA-level adaptability, and improved security, and reliability.

%A traditional method to bridge this performance adaptability divide has been the use of Engineering Change Orders (ECOs), which allow limited post-fab modifications without requiring full-chip redesigns~\cite{ho2010eco, jiang2020engineering}. These methods support localized updates, including combinational and sequential logic rectification and timing closure enhancements. However, traditional ECO approaches are inherently limited in scope—they are primarily designed for static, one-time fixes and lack the capability to address evolving security threats or dynamic application requirements. Moreover, ASIC supply chains become increasingly vulnerable to attacks such as intellectual property (IP) piracy and hardware Trojans. The inability of ECOs to support security-aware or post-deployment modifications renders ASICs susceptible to further threats. Additionally, the rigid and pre-fabricated nature of ECO-based corrections limits their utility in enabling hardware adaptability, thereby constraining long-term sustainability and reusability in rapidly evolving domains.
Engineering Change Orders (ECOs) enable limited post-fab updates, such as, logic rectification and timing closure, without requiring full-chip redesigns~\cite{ho2010eco, jiang2020engineering}. However, they are inherently static, localized, intended for one-time fixes, and cannot accommodate evolving application demands or emerging security threats. As ASIC supply chains grow more vulnerable to IP piracy and hardware Trojans~\cite{tehranipoor2024advances}, the lack of support for in-field or security-aware modifications leaves systems exposed. This rigidity ultimately hinders hardware adaptability, undermining long-term sustainability and reuse in fast-evolving domains.

\begin{wrapfigure}{r}{0.26\textwidth}
\centering
\vspace{-18pt}
\includegraphics[width=0.26\textwidth]{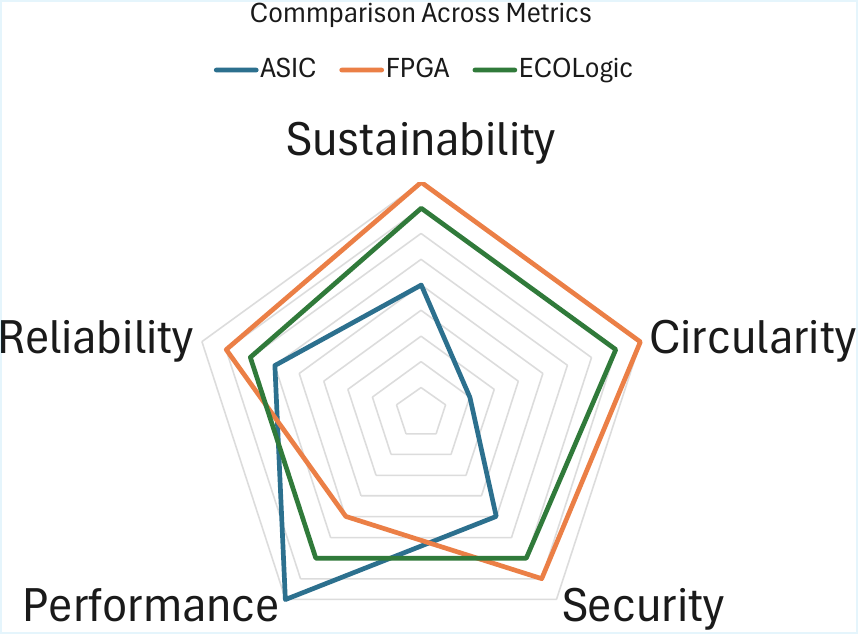}
\caption{Comparison of ASIC, FPGA, and \emph{ECOLogic} (ours) across five metrics.}
\vspace{-12pt}
\label{fig:asic_fpga_efpga}
\end{wrapfigure}
%To address the challenges of rigidity, obsolescence, and security vulnerability in conventional hardware design, in this paper we propose \emph{ECOLogic}, a hybrid System-on-Chip (SoC) architecture that integrates embedded FPGA (eFPGA) fabrics within an ASIC framework. \emph{ECOLogic} is inspired by the philosophy of engineering change, extending traditional ECO practices into the post-fab domain through dynamic reconfiguration. Simultaneously, it redefines the traditional notion of post-fab flexibility by promoting a new paradigm—hardware systems that are environmentally circular, obfuscation-capable, and dynamically adaptive. By selectively mapping volatile or security-sensitive IPs to reconfigurable logic, \emph{ECOLogic} supports in-field updates, logic redaction, and design reuse—significantly extending hardware lifespan while maintaining high performance and system reliability. Fig.~\ref{fig:asic_fpga_efpga} shows a comparison of ASIC, FPGA, and hybrid \emph{ECOLogic} across sustainability, circularity, security, performance, and reliability. This architecture leverages circular computing principles such as reuse, repurposing, and reconfiguration to reduce waste, while enhancing resilience to security threats and aging effects~\cite{rahman2024road}. The contributions of this work are summarized below.
To address the rigidity, obsolescence, and security challenges of conventional hardware, this paper proposes \emph{ECOLogic}, a hybrid System-on-Chip (SoC) architecture that embeds eFPGA fabrics within an ASIC framework. Inspired by the principles of engineering change, \emph{ECOLogic} extends traditional ECO practices into the post-fab domain through dynamic reconfiguration. It introduces a new paradigm of hardware design that economically circular, obfuscation-capable, and dynamically adaptive. By mapping volatile or security-sensitive IPs to reconfigurable eFPGA, \emph{ECOLogic} enables in-field updates, logic redaction, and device reuse, prolonging hardware lifespan while preserving high performance and reliability. Fig.\ref{fig:asic_fpga_efpga} compares ASIC, FPGA, and hybrid \emph{ECOLogic} in terms of sustainability, circularity, security, performance, and reliability. The architecture applies circular computing principles - reuse, repurposing, and reconfiguration to reduce waste and improve resilience against security threats and aging effects\cite{rahman2024road}. Our contributions are as follows:
\vspace{-3pt}
\begin{itemize}[leftmargin=*]
    \item \textsc{ECOLogic Architecture:} We propose \emph{ECOLogic}, a hybrid SoC framework that embeds reconfigurable logic within ASICs to enable hardware reuse, repurposing, and remanufacture—extending system lifetime (Sec.~\ref{sec:methodology}).

    \item \textsc{ECOScore Framework:} We introduce \emph{ECOScore}, a four-dimensional metric that guides selective eFPGA mapping based on adaptability, piracy threat, performance tolerance, and area fit. Our proposed scoring systematically shows up to $3.3$$\times$ higher \emph{ECOScore} for security and AI cores compared to static SoC components (Sec.~\ref{subsec:ecoscore_analysis}).

    \item \textsc{Security Enhancement:} By relocating sensitive IPs into reconfigurable fabric, \emph{ECOLogic} reduces the attack surface and supports secure post-fab updates, outperforming ASIC-only SoCs in confidentiality and exposure (Sec.~\ref{sec:security}).

    \item \textsc{Aging and Thermal Mitigation:} We demonstrate that \emph{ECOLogic} enables adaptive logic remapping under thermal and aging stress, maintaining up to $93.69\%$ improvement in timing slack at 140\textdegree{}C~(Sec.~\ref{subsec:age}).

    \item \textsc{Sustainability Impact:} \emph{ECOLogic} achieves up to 99$\times$ lower deployment carbon footprint than FPGA implementations by avoiding full-chip re-spins and enabling partial reconfigurability (Sec.~\ref{sub:sus_analy}).
\end{itemize}

\setlength{\arrayrulewidth}{0.6pt} % Darker vertical lines
\begin{table}[t]
\centering
\caption{The motivation matrix showing limitations of existing platforms and the need for a flexible, reliable, and secure hybrid SoC architecture.}
\label{tab:motivation_matrix}
\renewcommand{\arraystretch}{1.2}
\setlength\tabcolsep{4pt}
\begin{tabular}{c|c|c|c}
\textbf{Challenges} & \textbf{ASIC} & \textbf{FPGA} & \textbf{\emph{ECOLogic}} \\
\Xhline{1pt}
\makecell{Design \\ Updates} & 
\makecell{Limited \\ ECO} & 
\makecell{Flexible but \\ high overhead} & 
\makecell{Update only \\ selective logic} \\
\Xhline{1pt}
\makecell{IP \\ Protection} & 
\makecell{Vulnerable to\\ untrusted foundry} & 
\makecell{Weak bitstream \\ protection} & 
\makecell{Redacted and \\ hidden logic} \\
\Xhline{1pt}
Obsolescence & 
\makecell{Frequent \\ re-spins} & 
\makecell{Reusable with \\ performance hit} & 
\makecell{Reuse with \\ partial edits} \\
\Xhline{1pt}
\makecell{Security \\ Policy Update} & 
\makecell{No post-fab \\ patching} & 
\makecell{Coarse-grain \\ updateable} & 
\makecell{Runtime \\ patching} \\
\Xhline{1pt}
\makecell{Reliability} & 
Fails silently & 
Radiation-prone & 
\makecell{Can recover \\ by reconfig} \\
\Xhline{1pt}
\makecell{Area/Power \\ Efficiency} & 
\makecell{High perf., \\ fixed logic} & 
\makecell{Heavy logic \\ overhead} & 
\makecell{Overhead only \\ where needed} \\
\Xhline{1pt}
\makecell{Carbon \\ Footprint} & 
\makecell{High due to \\ re-fabrication} & 
\makecell{High due to \\ area waste} & 
\makecell{Lower via \\ reuse} \\
\\
\end{tabular}
\end{table}
\setlength{\arrayrulewidth}{0.4pt} % Darker vertical lines

\noindent The remainder of this paper is organized as follows. Sec.~\ref{sec:background} reviews the relevant background on ASIC-FPGA integration and outlines the limitations. Sec.~\ref{sec:methodology} presents the \emph{ECOLogic} architecture, detailing its design methodology, reconfigurable fabric integration, and the proposed \emph{ECOScore} IP selection framework. Sec.~\ref{sec:security} performs security, sustainability, and circularity analysis of \emph{ECOLogic}. Sec.~\ref{sec:result} presents a comprehensive evaluation of performance, reliability, and sustainability across six benchmark designs. Finally, Sec.~\ref{sec:conclusion} concludes the paper with future research directions.

% Paper Organization: Briefly summarize the remaining sections (e.g., methodology, implementation, experimental results, discussion, and conclusion).

\section{Background and Motivation}\label{sec:background}
\noindent This section outlines ASIC and FPGA-based architectural limitations and motivates the need for a hybrid solution that integrates post-fab flexibility with long-term resilience.
% 1. Why does this problem matter? What are the trends in AI, HPC, Defense, Automotive, and Space sectors?

% 2. How are these problems solved with existing technology? What are their drawbacks? 

% 3. Start with a motivating example where it can be beneficial to use eFPGA-integrated ASIC. How would that solve the drawback we just discussed? 

\subsection{Challenges with Modern Microelectronics}
\noindent ASIC and FPGA face critical challenges in sustainability, security, and performance due to their architecture and constraints.

\textsc{\underline{Sustainability:}} Sustainability challenges stem from the semiconductor industry’s carbon footprint (CFP), which includes embodied emissions from manufacturing, disposal, and recycling, as well as operational emissions from energy use~\cite{sudarshan2024eco}. ASICs exacerbate e-waste due to their rigid designs requiring carbon-intensive re-spins for updates, leading to rapid obsolescence ($5–8$ year lifespan)~\cite{sudarshan2023greenfpga}. Their fixed structure results in frequent replacements, increasing CFP~\cite{sudarshan2023greenfpga}. FPGAs offer partial mitigation through reconfigurability, enabling reuse across applications and reducing premature disposal~\cite{sudarshan2023greenfpga}. However, their larger area and higher power consumption elevate initial embodied and operational CFP~\cite{farooq2012fpga}. Circular economy principles, such as reuse and repurposing, are critical to minimizing lifecycle impacts and achieving decarbonization~\cite{doe2024circularity}.

\textsc{\underline{Security:}} Security vulnerabilities like reverse engineering, IP theft, and hardware trojans arise from reliance on untrusted third-party foundries, driven by the consolidation of costly fabrication facilities~\cite{rahman2020defense}. post-fab security flaws necessitate costly redesign of ASICs due to their static structure and lack of adaptability leaving them exposed to emerging threats. Beerel et. al.~\cite{beerel2022towards} formally proved that only universal circuits (reconfigurable) can preserve IP confidentiality with limited overhead. FPGAs provide limited security advantages through reconfigurable logic and encryption (e.g., bitstream encryption, IP watermarking)~\cite{zhang2019recent}, however their performance inefficiencies hinder adoption in high-performance applications.% this last line can be removed if needed.

\textsc{\underline{Performance:}} The performance-flexibility trade-off remains central to microsystems. ASICs deliver unmatched computational and power efficiency for fixed tasks but lack post-fab adaptability. Conversely, FPGAs offer reconfigurability for evolving applications, such as adaptive edge computing, but sacrifice efficiency due to their universal circuitry. This dichotomy creates inefficiencies - ASICs become obsolete as workloads evolve, while FPGAs underperform in power-constrained environments. Bridging this gap requires architectures that integrate ASIC-like efficiency with FPGA-like flexibility, enabling sustainable, secure, and high-performance computing for next-generation applications.

\subsection{Related Work}
\noindent Recent research has explored hybrid ASIC-eFPGA architectures to improve flexibility, performance, and security. Arnold introduces a RISC-V SoC for IoT end-nodes, integrating an eFPGA to accelerate analytics and cryptographic tasks with tight coupling and low-leakage design~\cite{schiavone2021arnold}. FlexBex embeds eFPGA fabric directly into a RISC-V CPU pipeline, enabling dynamic instruction set customization via partial reconfiguration~\cite{dao2020flexbex}. Abideen et al. propose a security-focused hASIC flow, using reconfigurable LUTs for obfuscation and IP protection within standard ASIC synthesis flows~\cite{abideen2023security}.
While prior works advance flexibility, performance, and security, none address the holistic design trade-offs involving sustainability and circularity in hybrid ASIC-eFPGA systems. GreenFPGA analyzes the CFP of traditional ASIC and FPGA platforms but does not evaluate integrated hybrid solutions~\cite{sudarshan2023greenfpga}. Similarly, ECO-CHIP focuses on chiplet-based architectures without considering the environmental or lifecycle implications of reconfigurable logic within monolithic SoCs~\cite{sudarshan2024eco}. To the best of our knowledge, this is the first work to jointly explore the sustainability, circularity, and security dimensions of a hybrid ASIC-eFPGA architecture. By unifying these domains, ECOLogic demonstrates how post-fab configurability enables lifecycle extension, in-field adaptability, and robust IP protection, establishing a new paradigm for security and sustainability.

\subsection{Motivation}

\noindent Traditional functional ECOs in ASICs use pre-placed spare cells to implement late-stage bug fixes or timing adjustments without full re-synthesis or re-layout~\cite{chang2008reap, ho2010eco}. However, they are limited to predefined regions and often induce routing congestion or timing violations when changes exceed available capacity. Metal-configurable gate-array cells offer slightly greater flexibility~\cite{chang2013eco}, but remain inadequate for complex or frequent updates. In contrast, FPGAs enable full post-fab reconfigurability, making them attractive for runtime adaptability—but at a steep cost. FPGAs suffer from high area and power overheads and significantly higher carbon footprints due to inefficient silicon use~\cite{sudarshan2023greenfpga}. This creates a fundamental trade-off: ASICs offer fixed, efficient performance, while FPGAs provide flexibility with poor efficiency. This dichotomy is especially pronounced in mission-critical domains like space, defense, and automotive systems. FPGAs support in-field updates and secure workloads~\cite{boada2011trends, montealegre2015flight, rahman2024road}, but are power-hungry and radiation-prone. ASICs are optimal for fixed high-performance tasks like radar or encryption~\cite{skup2015mixed, shim2024survey}, but require costly re-spins for even minor changes. To bridge this gap, embedding eFPGAs within ASIC SoCs has emerged as a compelling approach~\cite{schiavone2021arnold}. This hybrid model enables selective post-fab reconfiguration, combining ASIC-grade efficiency with FPGA-level flexibility. By supporting dynamic updates without full-chip redesign, eFPGA-augmented architectures overcome limitations of both extremes. Table~\ref{tab:motivation_matrix} summarizes these trade-offs and motivates the \emph{ECOLogic} framework.

\begin{figure}[!t]
\centering
\includegraphics[width=1.0\linewidth]{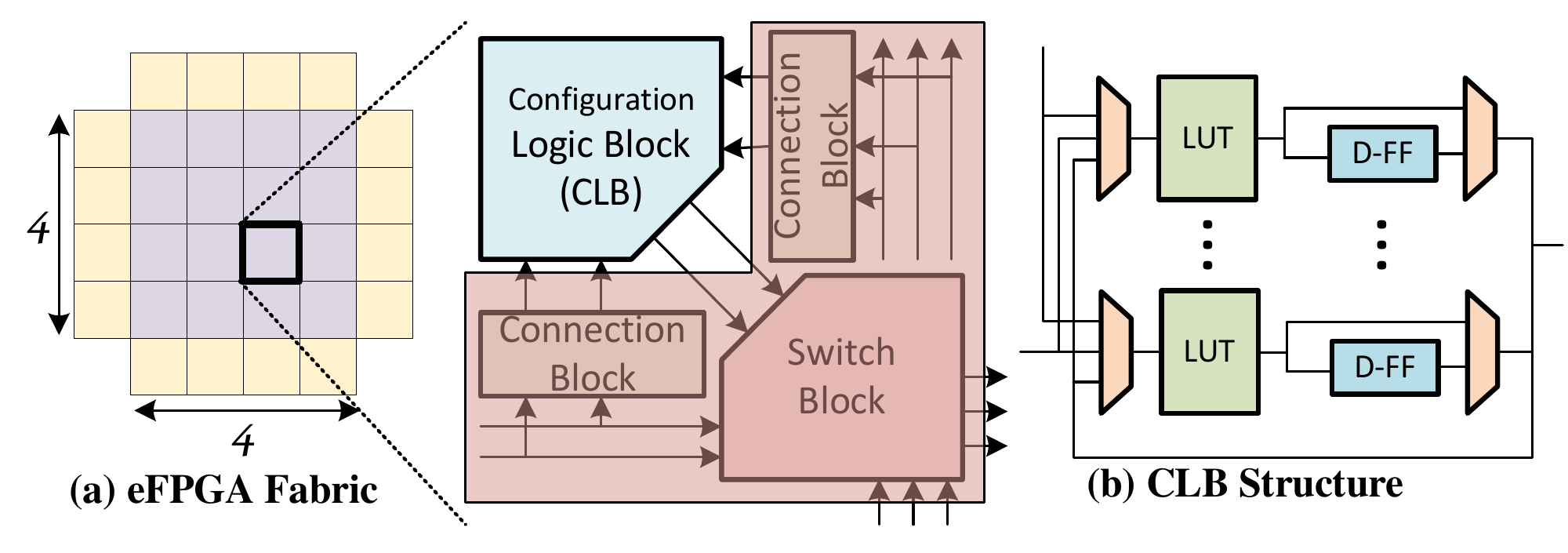}
\caption{eFPGA fabric architecture and its fundamental components. (a) A conceptual illustration of a 
$4\times4$ eFPGA fabric, highlighting the internal elements of each tile. (b) A detailed view of the Configuration Logic Block (CLB) within a tile, comprising LUTs, FFs, and multiplexers.}
\label{fig:efpga_overview}
\end{figure}

\section{Methodology}\label{sec:methodology}

\noindent This section details the design and implementation of \emph{ECOLogic}. We motivate selective reconfiguration through case studies, describe the physical architecture and fabric integration, introduce a scoring framework for IP selection, and outline the full design flow from synthesis to system integration.

\vspace{-3pt}
\subsection{Case Study: Evolving SoC Paradigm}\label{subsec:case_study}

\noindent Modern SoC architectures emphasize modularity and integration, combining fixed-function subsystems and specialized accelerators to meet performance and application require~\cite{tasnia2025opl4gpt,tasnia2025veriopt}. For instance, the Ballast SoC integrates high/medium-performance CPUs, system control, artificial intelligence (AI) accelerators, ethernet, and DSP engines~\cite{rautakoura2022ballast}, while a mixed-signal RISC-V design includes a DSP accelerator, fourier transform engine, and control IPs~\cite{bailey2019mixed}. Across generations, many legacy IPs, such as, CPUs, interconnects, and control units remain stable and unchanged~\cite{saiham2025leveraging, farahmandi2023cad}, whereas domain-specific AI and signal processing engines evolve rapidly in response to continuous innovation. This heterogeneous update pattern highlights a growing inefficiency in traditional monolithic ASIC design, where small changes to select IPs often require costly and carbon-intensive re-spins. \uline{To address this, \emph{ECOLogic} adopts a hybrid integration strategy - static backbone IPs are hardened in ASIC, while volatile or security-sensitive modules are implemented using embedded eFPGA fabric}. This allows partial reconfiguration of the targeted IPs, eliminating unnecessary redesign, improving system adaptability, and enabling reuse rather than replacement.
\vspace{-4pt}

\subsection{ECOLogic Architecture Overview}
\noindent \emph{ECOLogic} is a hybrid SoC architecture that integrates eFPGA fabric alongside hardened ASIC. As illustrated in Fig.~\ref{fig:efpga_overview}, the eFPGA region is composed of configurable logic blocks (CLBs), including look-up tables (LUTs), flip-flops (FFs), and multiplexers, organized in a tile-based architecture. These tiles can be reprogrammed to support post-fab design modifications or security updates. As highlighted in Sec.~\ref{subsec:case_study}, \emph{ECOLogic} maps stable, frequently reused IPs, such as, control units, interconnects, and general-purpose cores to fixed-function ASIC, and allocates volatile or security-sensitive IPs to the eFPGA. %This enables localized hardware updates without necessitating a full-chip redesign, improving system adaptability. 
While eFPGA integration introduces additional complexity and overhead compared to traditional ASIC, it supports hardware redaction, post-fab patching, and lifecycle extension, aligning with circular design principles~\cite{mohan2021hardware}.
\vspace{-6pt}
\subsection{System Integration and Fabric Configuration}

\noindent The internal communication in \emph{ECOLogic} adopts a hierarchical bus-based interconnect, with localized switch and connection blocks for efficient intra-fabric routing. Configuration is bitstream-driven, supporting both frame-based~\cite{koch2021fabulous} and scan-chain~\cite{mohan2021top} loading for runtime flexibility. The fabric connects to the ASIC domain via AXI protocols, allowing memory-mapped control of reconfigurable regions. On the ASIC side, static components—processors, controllers, and memory subsystems—are hardened at the technology node to ensure high performance and energy efficiency. The eFPGA fabric is synthesized using open-source tools like \textsc{Yosys}~\cite{wolf2013yosys} and \textsc{FABulous}~\cite{koch2021fabulous}, enabling custom configurations tailored to each IP. This flow supports post-fabrication updates for AI inference, cryptographic logic, and more, while remaining scalable across domains. Though it incurs some area and power overhead, this integration substantially extends hardware lifetime and adaptability.

\vspace{-6pt}

\subsection{ECOScore: IP Selection Guideline in \emph{ECOLogic}}\label{sec:ecoscore}

\noindent To guide RTL partitioning in \emph{ECOLogic}, we introduce \emph{ECOScore}, a quantitative framework that ranks candidate IPs across four dimensions: \textit{adaptability}, \textit{piracy threat}, \textit{performance tolerance}, and \textit{resource fit}. These reflect key design concerns—frequent updates, IP vulnerability, timing resilience, and reconfigurability. Scores from RTL and synthesis metrics are combined to enable tunable, scalable redaction.

\textsc{Adaptability} ($A_i$) captures how actively the RTL for IP $i$ evolves across development cycles. Frequent changes justify using reconfigurable fabric to avoid full re-spins. We define the score as:
\vspace{-6pt}
\begin{equation}
A_i = \frac{\ln(1 + \mathrm{LOC\_changed}_i)}{\ln(1 + \max_j \mathrm{LOC\_changed}_j)}
\label{eq:adaptability}
\end{equation}
\noindent where $\mathrm{LOC\_changed}_i$ denotes the line-of-code changes for IP $i$. The logarithmic normalization tempers extreme differences and avoids biasing against smaller blocks that naturally undergo less churn.

\textsc{Piracy Threat} ($O_i$) evaluates how critical it is to redact a given IP block from the main ASIC fabric to prevent reverse engineering or overproduction. It considers three sub-metrics: designer-assigned confidentiality risk ($C_i$), exposure factor ($E_i$), and redaction potential ($R_i$). Specifically:
\vspace{-6pt}
\begin{align}
E_i &= \frac{\# \text{I/O and control nets}}{\# \text{internal nets + state elements}} \label{eq:exposure} \\
R_i &= \frac{\text{logic mapped to eFPGA}}{\text{total logic in IP}} \nonumber
\end{align}
\noindent $E_i$ estimates attack surface, while $R_i$ reflects how much of the logic can feasibly be redacted. The final threat score is computed as:
\vspace{-6pt}
\begin{equation}
O_i = \mu C_i + \nu E_i + \xi R_i,\quad \mu + \nu + \xi = 1
\label{eq:obfuscation}
\end{equation}
\noindent where $(\mu, \nu, \xi)$ are designer-specified weights to balance design sensitivity, surface exposure, and redaction coverage. A high $O_i$ suggests the IP is security-critical and benefits from obfuscation via reconfigurable mapping.

\textsc{Performance Tolerance} ($P_i$) quantifies whether an IP can tolerate slower execution due to eFPGA-induced delay. It is based on the ratio between ASIC and eFPGA maximum operating frequencies:
\vspace{-6pt}
\begin{equation}
P_i = 1 - \min\left(1, \frac{F^{\text{ASIC}}_{\max,i} - F^{\text{eFPGA}}_{\max,i}}{F^{\text{ASIC}}_{\max,i}} \right)
\label{eq:performance}
\end{equation}
\noindent where $F^{\text{ASIC}}_{\max,i}$ and $F^{\text{eFPGA}}_{\max,i}$ denote the synthesized clock speeds. The closer these values are, the higher the $P_i$, indicating minimal performance loss and thus high suitability for reconfiguration.

\begin{figure}[!t]
\centering
\includegraphics[width=1.0\linewidth]{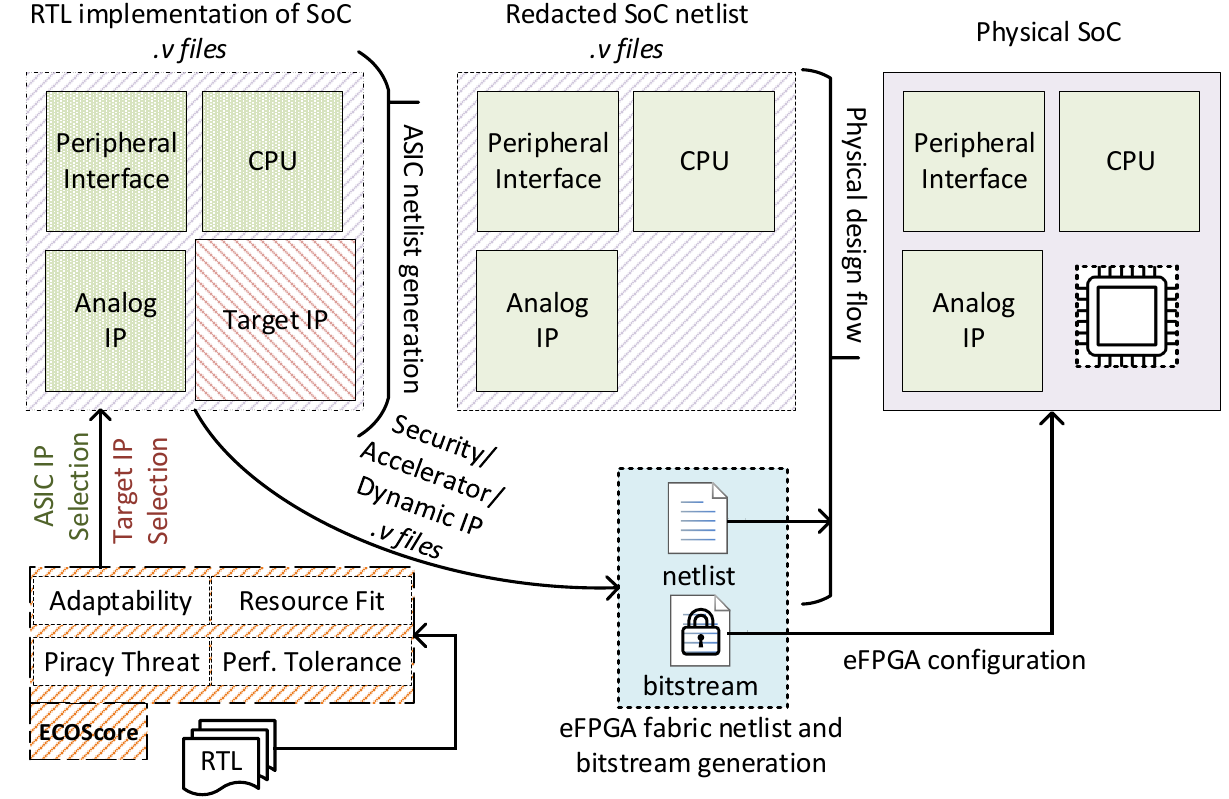}
\caption{\emph{ECOLogic} design flow for sustainable computing illustrating the RTL implementation, selection of security, accelerator or dynamic  IP for eFPGA configuration, and fabricating to physical SoC.}
\label{fig:flow}
\end{figure}

\textsc{Resource Fit} ($R_i$) considers the area required to place an IP within the constrained eFPGA fabric. Let $A_i$ be the total synthesized area (in gates or $\mu$m\textsuperscript{2}). We normalize across all IPs to capture their relative footprint:

\vspace{-6pt}
\begin{equation}
R_i = \frac{A_{\max} - A_i}{A_{\max} - A_{\min}}
\label{eq:resource_fit}
\end{equation}
\noindent Here, $A_{\max}$ and $A_{\min}$ are the largest and smallest observed IP areas in the design set. A higher $R_i$ indicates lower area overhead, reducing placement congestion and enabling more flexible dynamic updates.

\noindent Finally, we define the total score as a convex combination of the four normalized metrics:
\vspace{-6pt}
\begin{equation}
\text{ECOScore}_i = \alpha A_i + \beta O_i + \gamma P_i + \delta R_i,\quad \alpha + \beta + \gamma + \delta = 1
\label{eq:ecoscore}
\end{equation}
\noindent The weights $(\alpha, \beta, \gamma, \delta)$ reflect design intent—e.g., prioritizing security, performance, or fabric utilization. This formulation enables principled RTL partitioning that balances reconfigurability against physical and functional constraints.

\vspace{-0.1in}

\subsection{ECOLogic Design Flow}

% Explain for the FlexCore is integrated in to the whole ASIC design flow. How the rest of the ASIC is designed/implemented, how the eFPGAs are configured

\noindent Once a candidate IP is selected using the \emph{ECOScore} framework, the \emph{ECOLogic} flow integrates it into an ASIC augmented with an embedded eFPGA fabric. This hybrid design supports selective reconfigurability while maintaining ASIC-level efficiency and performance. As shown in Fig.~\ref{fig:flow}, the IP is first described in HDL and synthesized using \textsc{Yosys}~\cite{wolf2013yosys} to generate netlists and resource estimates (LUTs, FFs, DSPs). These metrics guide \textsc{FABulous}~\cite{koch2021fabulous} in mapping the logic to a tile-based eFPGA layout. The completed fabric is integrated into the ASIC using commercial tools such as \textsc{Synopsys Design Compiler}, ensuring compatibility with standard physical design, timing, and power flows.

% This approach allows \emph{ECOLogic} to support post-fabrication updates, late-stage patching, and lifecycle extension for selected IPs—all without requiring costly full-chip re-spins.

%The hybrid architecture of ASIC and eFPGA offers a balance between computational efficiency and adaptability. The Verilog-to-bitstream flow of \texttt{FABulous} enables dynamic post-fabrication reconfiguration of the eFPGA portion, facilitating real-time functional updates without hardware replacement. 
Unlike traditional ASICs that demand costly and time-consuming re-spins for even minor design updates, \emph{ECOLogic} enables in-field reconfiguration of selected IPs, thereby reducing electronic waste and lowering the overall CFP. By promoting hardware longevity, sustainability, and security through selective reprogrammability, \emph{ECOLogic} defines a new paradigm for environmentally sustainable semiconductor design.

\vspace{-0.1in}

\section{Security, Sustainability, and Circularity Analysis of \emph{ECOLogic}}\label{sec:security}
% Describe various security attacks, and how eFPGA integrated ASIC can counteract them. 
\noindent In this section we challenge \emph{ECOLogic} against various security threats in modern semiconductor supply chain and analyze its resiliency. Additionally, we discuss the sustainability and circularity benefits of eFPGA in extending the SoC lifecycle.

\subsection{Security Analysis}\label{subsec:sec_analysis}

\noindent Integrating eFPGAs within ASICs offers versatile security enhancements by enabling post-fabrication configurability. We categorize the security benefits of this integration below.

\subsubsection{IP Protection via eFPGA Redaction}
eFPGA-based redaction conceals critical IP by replacing sensitive modules with programmable fabric, which remains inactive until configured via a secure bitstream~\cite{Bhandari2021}. This approach thwarts IP theft, reverse engineering, and overproduction by ensuring that the full design functionality is revealed only after deployment by trusted parties. Additionally, the complexity of custom eFPGA fabrics increases resilience against de-obfuscation.

\subsubsection{Dynamic Security Monitoring and Policy Enforcement}
Reconfigurable logic within eFPGAs enables in-field updates of hardware-based security monitors and policies. This dynamic capability allows to adapt to new threats over time without needing redesign, supporting runtime enforcement of integrity, privilege separation, and access controls~\cite{rahman2024road}.

\begin{figure}[t]
    \centering
    \includegraphics[width=1.0\linewidth]{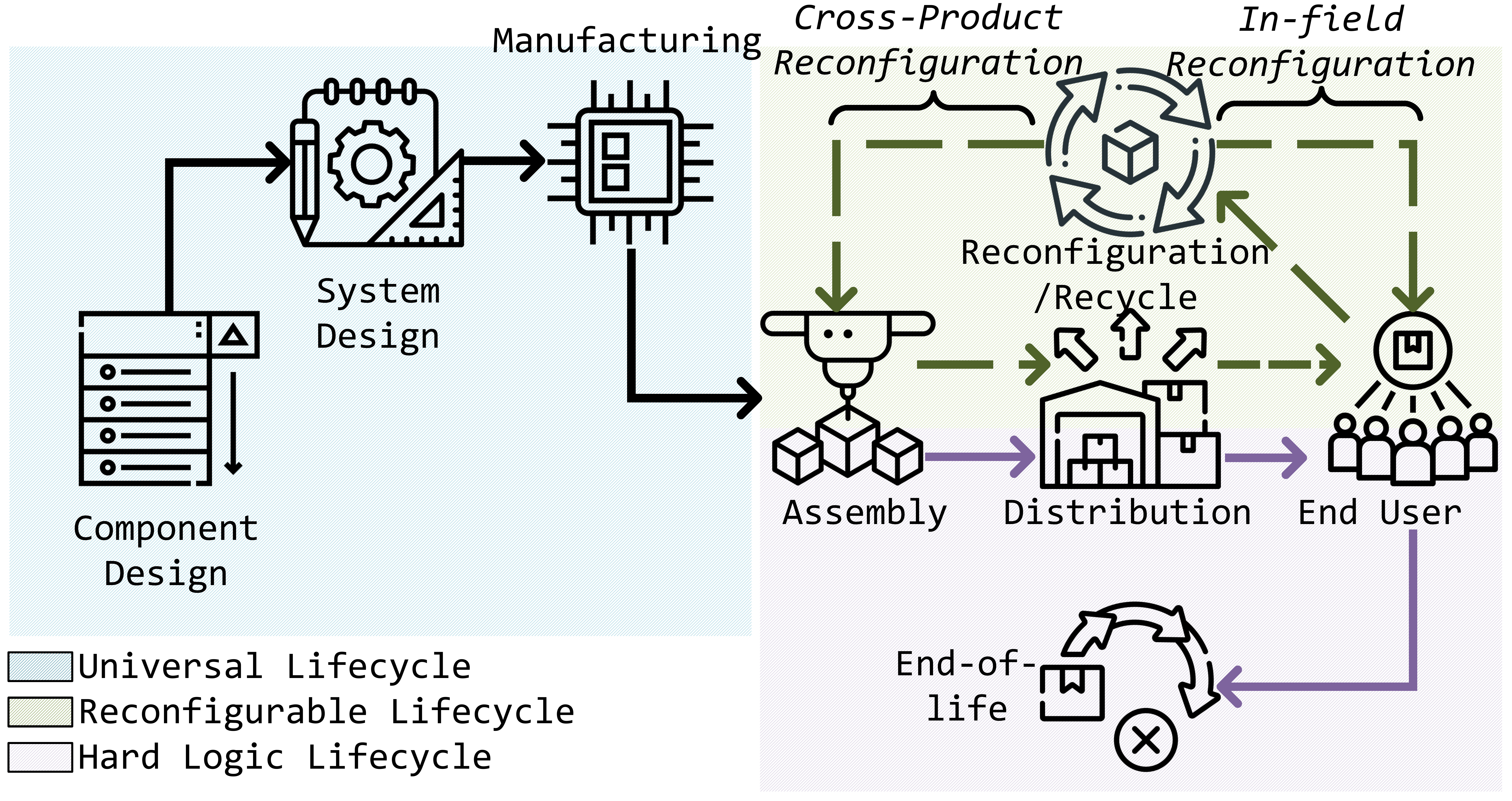}
    \caption{Lifecycle flow of semiconductor systems across three design paradigms. \emph{ECOLogic} uniquely enables circularity via in-field and cross-product reconfiguration, allowing components to be reused, repaired, or repurposed after deployment—capabilities not available in traditional ASIC.}
    \label{fig:semiconductor_LCA}
\end{figure}

\subsubsection{Side-Channel Monitoring and Leakage Mitigation}
eFPGAs can be used to embed side-channel-aware monitors, such as TrustGuard~\cite{zhang2024trustguard}, directly into the chip. These monitors analyze power traces to detect malware, code injection, and reuse attacks in real-time. Their integration reduces reliance on external sensors and enhances on-chip anomaly detection with low latency and high accuracy.

\subsubsection{Resilience to Evolving Attacks}
The runtime reconfigurability and inherent architectural complexity of eFPGAs create a moving target for attackers. Cyclic routing, large bitstream key spaces, and dynamic logic updates reduce fixed attack surfaces and hinder RE and SAT-based attacks~\cite{Rezaei2022, Bhandari2021}.

\vspace{-0.1in}
\subsection{Sustainability Analysis}\label{sub:sus_analy}

\noindent \emph{ECOLogic} promotes sustainable chip design by enabling selective post-fabrication reconfiguration of IP blocks, reducing the need for re-manufacturing and extending hardware lifespan. Unlike traditional ASICs that require costly full-chip re-spins for updates, \emph{ECOLogic} facilitates hardware reuse across deployments, thereby lowering carbon emissions.

As shown in Fig.~\ref{fig:semiconductor_LCA}, \emph{ECOLogic} supports both in-field and cross-product reconfiguration. This allows chips to be repurposed, updated, or recycled after initial deployment—capabilities not available in fixed-function ASICs. Such reuse eliminates the manufacturing carbon footprint for each new application.

To quantify the environmental benefit, we model the total CFP of \emph{ECOLogic} across $N_{\text{app}}$ applications using GreenFPGA~\cite{sudarshan2023greenfpga}. The total footprint includes only deployment-phase emissions, as the hardware is manufactured once:

\vspace{-0.2in}
\begin{equation} \label{total-CFP}
C_{\text{\textsc{ECOLogic}}} = \sum_{i=1}^{N_{\text{app}}} \left( T_i \cdot C_{\text{deploy},i} \right)
\end{equation}
\vspace{-0.2in}

where $T_i$ is the runtime duration of application $i$, and $C_{\text{deploy},i}$ is its corresponding deployment-phase CFP.

Deployment carbon is further broken down into energy usage during runtime and application development:

\begin{equation}\label{deploy carbon}
C_{\text{deploy}} = N_{\text{vol}} \cdot C_{\text{src,use}} \cdot E_{\text{use}} + C_{\text{app-dev}},
\end{equation}

where $N_{\text{vol}}$ denotes the number of deployed instances, $E_{\text{use}}$ is the runtime energy per application, and $C_{\text{src,use}}$ represents the grid's carbon intensity. $C_{\text{app-dev}}$ includes CPU energy from RTL/HLS synthesis and bitstream generation~\cite{sudarshan2024eco}.Table~\ref{tab:deploy-params} summarizes all parameters used in the calculation.

\setlength{\arrayrulewidth}{0.6pt} % Darker vertical lines
\begin{table}[t]
\centering
\caption{Deployment-phase parameters used for carbon footprint estimation.}
\label{tab:deploy-params}
\renewcommand{\arraystretch}{1.2}
\setlength\tabcolsep{4pt}
\begin{tabular}{>{\centering\arraybackslash}p{0.20\columnwidth}|
                >{\centering\arraybackslash}p{0.32\columnwidth}|
                >{\centering\arraybackslash}p{0.18\columnwidth}|
                >{\centering\arraybackslash}p{0.14\columnwidth}}
\textbf{Parameter} & \textbf{Description} & \textbf{Units} & \textbf{Value} \\
\hline
$N_{\rm vol}$ & Application deployment volume & \# & $1 \times 10^6$ \\
\hline
$T$ & Runtime per application & hours & 131{,}400 \\
\hline
$N_{\rm app}$ & Number of applications mapped & \# & 5 \\
\hline
$E_{\rm use}$ & Runtime energy per application & kWh & derived~\cite{sudarshan2024eco} \\
\hline
$C_{\rm src,use}$ & Grid carbon intensity & kg CO$_2$/kWh & 700 \\
\hline
CPU power/core & Synthesis/configuration power per core & W & 10 \\
\hline
CPU cores & Total cores used & \# & 8 \\
\hline
RTL synthesis time & RTL front-end time per app & hours & 2.5 \\
\hline
HLS synthesis time & Backend HLS time per app & hours & 1.0 \\
\hline
Config time & Bitstream reconfiguration time & hours & 0 \\
\hline
$C_{\text{app-dev}}$ & App development carbon cost & kg CO$_2$ & derived \\
\end{tabular}
\end{table}
\setlength{\arrayrulewidth}{0.4pt} % Reset to default

\subsection{Circularity Analysis}
\label{sub:cir_analy}

\noindent The \emph{ECOLogic} architecture reimagines semiconductor circularity by enabling hardware to adapt with evolving application demands, rather than becoming obsolete. As illustrated in Fig.~\ref{fig:semiconductor_LCA}, \emph{ECOLogic} diverges from the traditional linear ASIC lifecycle—design, manufacture, deploy, discard—by supporting remanufacture, in-field repair, and repurposing within the same hardware instance.

Unlike fixed-function ASICs, \emph{ECOLogic} enables localized updates through its embedded eFPGA fabric. When defects emerge or application needs shift, only the targeted reconfigurable modules are reprogrammed—avoiding costly re-spins or full-chip replacements. A single hardware instance can be reused across applications via partial reconfiguration, and IPs originally designed for cryptographic tasks can be repurposed for roles like AI inference. Faulty fabric regions can be bypassed or reconfigured to extend lifetime without new silicon. While material-level recycling is unchanged, \emph{ECOLogic} indirectly supports it by prolonging hardware utility and reducing production frequency. These features collectively support a circular model of use, minimizing carbon impact and improving system-level sustainability.

\section{Results} \label{sec:result}
\noindent This section validates the effectiveness of \emph{ECOLogic} across key evaluation metrics: sustainability, aging resilience, and performance. We analyze six representative SoC designs to measure CFP, reliability under stress, and hardware efficiency compared to traditional FPGAs and ASICs using \emph{ECOScore}, a scoring framework to guide reconfigurable IP partitioning.

\setlength{\arrayrulewidth}{0.6pt} % Darker vertical lines
\begin{table}[t]
\centering
\caption{Normalized \emph{ECOScore} components for six benchmark IPs, used to guide eFPGA mapping decisions in the \emph{ECOLogic} framework.}

\label{tab:ecoscore_matrix}
\renewcommand{\arraystretch}{1.2}
\setlength\tabcolsep{8pt}
\begin{tabular}{c|c|c|c|c|c}
\textsc{Design} & \makecell{$A_i$\\Adapt.} & \makecell{$O_i$\\Obfus.} & \makecell{$P_i$\\Perf.} & \makecell{$R_i$\\Resrc.} & \textsc{ECOScore}\\
\Xhline{1pt}
Design 1 & 0.98 & 1.00 & 0.88 & 0.47 & 0.86 \\
\Xhline{1pt}
Design 2 & 0.82 & 0.98 & 0.86 & 0.55 & 0.84 \\
\Xhline{1pt}
Design 3 & 1.00 & 0.82 & 0.75 & 0.00 & 0.72 \\
\Xhline{1pt}
Design 4 & 0.94 & 0.79 & 0.79 & 0.13 & 0.75 \\
\Xhline{1pt}
Design 5 & 0.19 & 0.25 & 0.97 & 1.00 & 0.64 \\
\Xhline{1pt}
Design 6 & 0.26 & 0.31 & 1.00 & 0.84 & 0.67 \\
\end{tabular}
\end{table}
\setlength{\arrayrulewidth}{0.4pt} % Reset to default

\subsection{Case Study for \emph{ECOScore} Evaluation}
\label{subsec:benchmark}

\noindent To demonstrate the utility of the \emph{ECOScore} framework, we evaluated six representative IP blocks for potential conversion to \emph{ECOLogic}. Designs 1 and 2 (ASCON and SHA-256) are cryptographic cores with high confidentiality and post-fab update needs. Designs 3 and 4 (Transformer and CNN) are AI accelerators known for frequent updates and adaptability, though limited by performance sensitivity and larger area. Designs 5 and 6 (CVA6 Interconnect and Controller) are SoC modules that benefit from in-field reconfigurability despite their relative stability. Each design was scored across adaptability, piracy threat, performance tolerance, and resource fit. While this study used fixed weights favoring security and flexibility, \emph{ECOScore} remains tunable—allowing designers to reweight based on set specific goals.

\subsection{ECOScore Analysis} \label{subsec:ecoscore_analysis}

\noindent This subsection analyzes the \emph{ECOScore} sub-metrics for each IP block, with the results summarized in Table~\ref{tab:ecoscore_matrix}.

\textsc{Adaptability ($A_i$)} measures log-scaled RTL churn across the last three design revisions, capturing how frequently each IP undergoes functional or structural updates. Design~3 (transformer accelerator) had the highest churn at approximately 200 LOC changes, yielding \(A_3 = 0.93\), indicating rapid algorithmic evolution and a strong case for reconfigurable deployment. Design~1 (ASCON) and Design~4 (CNN) followed closely with 180 and 175 LOC changed, giving \(A_1 = 0.91\) and \(A_4 = 0.87\), respectively—both are high-churn, domain-specific accelerators that benefit from post-fab flexibility. Design~2 (SHA-256) showed moderate churn (\(\sim\)150 LOC), resulting in \(A_2 = 0.76\), which aligns with its status as a cryptographic block with some evolving variants. In contrast, Design~6 (controller) and Design~5 (interconnect) experienced only minor updates (\(\sim\)30 and 15 LOC), scoring \(A_6 = 0.24\) and \(A_5 = 0.18\), respectively. These low scores suggest that these IPs are stable and infrequently modified—typical of control or routing logic hardened for reliability. This high $A_i$ implies the IP may require future patching or feature extensions, making it suitable for eFPGA mapping. Conversely, a low $A_i$ suggests stability, favoring static ASIC implementation.

\textsc{Piracy Threat ($O_i$)} evaluates the benefit of relocating an IP to reconfigurable logic for protection against IP theft, reverse engineering, and overproduction. The score is derived using the weighted sum \(O_i = 0.5\,C_i + 0.3\,E_i + 0.2\,R_i\), where $C_i$ is the confidentiality risk, $E_i$ is the I/O-to-internal net exposure ratio, and $R_i$ is the fraction of the IP mapped to the eFPGA fabric. Design~1 (ASCON) scored highest with \(O_1 = 1.00\), reflecting its highly sensitive cryptographic role, moderate exposure (\(E_1 \approx 0.25\)), and strong redaction coverage (\(R_1 = 0.80\)). Design~2 (SHA-256) also showed strong obfuscation potential with \(O_2 = 0.98\), driven by similar sensitivity and a slightly lower exposure. Designs~3 and 4 (transformer and CNN) achieved mid-range scores—\(O_3 = 0.82\), \(O_4 = 0.79\)—due to moderate sensitivity and higher exposure (e.g., large input/output interface) and partial mapping to eFPGA. Finally, the SoC logic in Designs~5 and 6 scored lowest (\(O_5 = 0.25\), \(O_6 = 0.31\)), consistent with their low confidentiality, minimal internal redaction, and relatively flat exposure profiles. A higher $O_i$ indicates stronger security gains when offloading to the fabric; a lower score implies that static logic is sufficient for threat modeling.

\textsc{Performance Tolerance ($P_i$)} quantifies how well an IP tolerates the slower clock speeds of eFPGA implementations. Control-intensive logic (Design~6, controller) achieved the highest retention with a minimal drop, yielding \(P_6 = 1.00\). The CVA6 interconnect (Design~5) also performed well with \(P_5 = 0.97\), owing to its moderate complexity and relatively flat frequency curve under reconfiguration. Cryptographic cores (Designs~1–2) maintained good timing characteristics despite redirection: \(P_1 = 0.88\) and \(P_2 = 0.86\). In contrast, the AI accelerators (Designs~3–4) experienced more substantial slowdowns due to deep pipelines and routing congestion: \(P_3 = 0.75\) and \(P_4 = 0.79\). Higher \(P_i\) values indicate more favorable frequency scaling and reduced performance penalties when mapped to reconfigurable logic, while lower scores suggest the need for careful partitioning or ASIC-only deployment for timing-critical workloads.

\begin{table}[t]
\centering
\caption{Deployment Carbon comparison between \emph{ECOLogic} and FPGA based on application lifetime.}
\label{tab:app_lifetime}
\resizebox{\columnwidth}{!}{%
\begin{tabular}{ll|llllllll}
\hline
\multicolumn{2}{c}{\multirow{3}{*}{Designs}} &
  \multicolumn{8}{c}{Deployment Cost ($\times 10^4$) (Eq. Kgs of $\mathrm{CO}_2$)} \\ \cline{3-10} 
\multicolumn{2}{c}{} &
  \multicolumn{4}{c|}{App. Lifetime (years)} &
  \multicolumn{4}{c}{App. Volume} \\ \cline{3-10} 
\multicolumn{2}{c}{} &
  \multicolumn{1}{c|}{0.2} &
  \multicolumn{1}{c|}{1} &
  \multicolumn{1}{c|}{2} &
  \multicolumn{1}{c|}{2.5} &
  \multicolumn{1}{c|}{$1\mathrm{K}$} &
  \multicolumn{1}{c|}{$6\mathrm{K}$} &
  \multicolumn{1}{c|}{$90\mathrm{K}$} &
  $1\mathrm{M}$ \\ \hline
\multicolumn{1}{l|}{\multirow{2}{*}{Design 1}} &
  \emph{ECOLogic} &
  \multicolumn{1}{c|}{0.93} &
  \multicolumn{1}{c|}{4.66} &
  \multicolumn{1}{c|}{9.32} &
  \multicolumn{1}{c|}{11.6} &
  \multicolumn{1}{c|}{0.00932} &
  \multicolumn{1}{c|}{0.0559} &
  \multicolumn{1}{c|}{0.839} &
  9.32 \\ \cline{2-10} 
\multicolumn{1}{l|}{} &
  FPGA &
  \multicolumn{1}{c|}{298} &
  \multicolumn{1}{c|}{$1.49\mathrm{K}$} &
  \multicolumn{1}{c|}{$2.98\mathrm{K}$} &
  \multicolumn{1}{c|}{$3.73\mathrm{K}$} &
  \multicolumn{1}{c|}{2.98} &
  \multicolumn{1}{c|}{17.9} &
  \multicolumn{1}{c|}{268} &
  $2.98\mathrm{K}$ \\ \hline
\multicolumn{1}{l|}{\multirow{2}{*}{Design 2}} &
  \emph{ECOLogic} &
  \multicolumn{1}{c|}{0.83} &
  \multicolumn{1}{c|}{4.17} &
  \multicolumn{1}{c|}{8.34} &
  \multicolumn{1}{c|}{10.4} &
  \multicolumn{1}{c|}{0.0083} &
  \multicolumn{1}{c|}{0.0501} &
  \multicolumn{1}{c|}{0.751} &
  8.34 \\ \cline{2-10} 
\multicolumn{1}{l|}{} &
  FPGA &
  \multicolumn{1}{c|}{273} &
  \multicolumn{1}{c|}{$1.37\mathrm{K}$} &
  \multicolumn{1}{c|}{$2.73\mathrm{K}$} &
  \multicolumn{1}{c|}{$3.42\mathrm{K}$} &
  \multicolumn{1}{c|}{2.73} &
  \multicolumn{1}{c|}{16.4} &
  \multicolumn{1}{c|}{246} &
  $2.73\mathrm{K}$ \\ \hline
\multicolumn{1}{l|}{\multirow{2}{*}{Design 3}} &
  \emph{ECOLogic} &
  \multicolumn{1}{c|}{0.878} &
  \multicolumn{1}{c|}{4.39} &
  \multicolumn{1}{c|}{8.78} &
  \multicolumn{1}{c|}{11} &
  \multicolumn{1}{c|}{0.0087} &
  \multicolumn{1}{c|}{0.052} &
  \multicolumn{1}{c|}{0.791} &
  8.78 \\ \cline{2-10} 
\multicolumn{1}{l|}{} &
  FPGA &
  \multicolumn{1}{c|}{286} &
  \multicolumn{1}{c|}{$1.43\mathrm{K}$} &
  \multicolumn{1}{c|}{$2.86\mathrm{K}$} &
  \multicolumn{1}{c|}{$3.57\mathrm{K}$} &
  \multicolumn{1}{c|}{2.86} &
  \multicolumn{1}{c|}{17.1} &
  \multicolumn{1}{c|}{257} &
  $2.86\mathrm{K}$ \\ \hline
\multicolumn{1}{l|}{\multirow{2}{*}{Design 4}} &
  \emph{ECOLogic} &
  \multicolumn{1}{c|}{0.754} &
  \multicolumn{1}{c|}{3.77} &
  \multicolumn{1}{c|}{7.54} &
  \multicolumn{1}{c|}{9.43} &
  \multicolumn{1}{c|}{0.00754} &
  \multicolumn{1}{c|}{0.045} &
  \multicolumn{1}{c|}{0.679} &
  7.54 \\ \cline{2-10} 
\multicolumn{1}{l|}{} &
  FPGA &
  \multicolumn{1}{c|}{286} &
  \multicolumn{1}{c|}{$1.43\mathrm{K}$} &
  \multicolumn{1}{c|}{$2.86\mathrm{K}$} &
  \multicolumn{1}{c|}{$3.57\mathrm{K}$} &
  \multicolumn{1}{c|}{2.86} &
  \multicolumn{1}{c|}{17.1} &
  \multicolumn{1}{c|}{257} &
  $2.86\mathrm{K}$ \\ \hline
\multicolumn{1}{l|}{\multirow{2}{*}{Design 5}} &
  \emph{ECOLogic} &
  \multicolumn{1}{c|}{647} &
  \multicolumn{1}{c|}{$3.24\mathrm{K}$} &
  \multicolumn{1}{c|}{$6.47\mathrm{K}$} &
  \multicolumn{1}{c|}{$8.09\mathrm{K}$} &
  \multicolumn{1}{c|}{6.47} &
  \multicolumn{1}{c|}{38.8} &
  \multicolumn{1}{c|}{583} &
  $6.47\mathrm{K}$ \\ \cline{2-10} 
\multicolumn{1}{l|}{} &
  FPGA &
  \multicolumn{1}{c|}{248} &
  \multicolumn{1}{c|}{$1.24\mathrm{K}$} &
  \multicolumn{1}{c|}{$2.48\mathrm{K}$} &
  \multicolumn{1}{c|}{$3.10\mathrm{K}$} &
  \multicolumn{1}{c|}{2.48} &
  \multicolumn{1}{c|}{14.9} &
  \multicolumn{1}{c|}{224} &
  $2.48\mathrm{K}$ \\ \hline
\multicolumn{1}{l|}{\multirow{2}{*}{Design 6}} &
  \emph{ECOLogic} &
  \multicolumn{1}{c|}{0.971} &
  \multicolumn{1}{c|}{4.86} &
  \multicolumn{1}{c|}{9.71} &
  \multicolumn{1}{c|}{12.1} &
  \multicolumn{1}{c|}{0.0097} &
  \multicolumn{1}{c|}{0.058} &
  \multicolumn{1}{c|}{0.874} &
  9.71 \\ \cline{2-10} 
\multicolumn{1}{l|}{} &
  FPGA &
  \multicolumn{1}{c|}{310} &
  \multicolumn{1}{c|}{$1.55\mathrm{K}$} &
  \multicolumn{1}{c|}{$3.10\mathrm{K}$} &
  \multicolumn{1}{c|}{$3.88\mathrm{K}$} &
  \multicolumn{1}{c|}{3.10} &
  \multicolumn{1}{c|}{18.6} &
  \multicolumn{1}{c|}{279} &
  $3.10\mathrm{K}$ \\ \hline
\end{tabular}%
}
\end{table}

\textsc{Resource Fit ($R_i$)} assesses how efficiently each IP maps onto the available eFPGA fabric by comparing synthesized area across all candidates. We compute a weighted area cost per IP and assign the highest score to the smallest area using Equation~\ref{eq:resource_fit}. Under this formulation, the CVA6 interconnect (Design~5) emerges as the most compact IP with \(R_5 = 1.00\), followed by the controller module (Design~6) at \(R_6 = 0.84\). These values reflect the low complexity and minimal logic depth of peripheral SoC logic. The cryptographic IPs—ASCON and SHA-256 (Designs~1 and 2)—occupy a mid-tier range with \(R_1 = 0.47\) and \(R_2 = 0.55\), balancing moderate area requirements against their security utility. On the lower end, the AI accelerators (Designs~3 and 4) consume the most area, yielding \(R_3 = 0.00\) and \(R_4 = 0.13\). These results signal that such blocks are least amenable to full eFPGA mapping unless fabric capacity is scaled or only partial components are offloaded.

\textsc{Composite \emph{ECOScore}} is calculated as a weighted sum of the four metrics using designer-defined weights \((\alpha, \beta, \gamma, \delta) = (0.25, 0.35, 0.20, 0.20)\). These values prioritize piracy threat and adaptability while accounting for performance tolerance and fit. Applying this weighting to the raw scores yields the following: Design~1 scores highest with \(0.86\), closely followed by Design~2 at \(0.84\), and Designs~3 and 4 at \(0.72\) and \(0.75\), respectively. The interconnect (Design~5) scores \(0.64\) and the controller (Design~6) scores \(0.67\). Normalizing these values against the highest (Design~1) produces final \emph{ECOScore} values of \(\{1.00,\,0.98,\,0.84,\,0.87,\,0.74,\,0.78\}\). These results indicate that cryptographic cores (Designs~1–2) are strong candidates for full eFPGA embedding. The interconnect (Design~5) also performs well in terms of footprint and performance tolerance but scores lower in security relevance. Meanwhile, the AI accelerators (Designs~3–4), though adaptable and obfuscation-friendly, face penalties from high area usage, suggesting that partial or hybrid reconfiguration being more appropriate.

\subsection{Carbon Footprint Analysis}
\label{subsec:carbon_footprint}

\noindent Table~\ref{tab:app_lifetime} presents the deployment carbon footprint (CFP) for each design, capturing emissions from application development and runtime operation. In five out of six cases, \emph{ECOLogic} significantly outperforms traditional FPGA-based implementations. For example, in Design~1, the $1$-year deployment cost drops from $1.49K$ to $4.66 \times10\textsuperscript{4}$ kg CO\textsubscript{2}, representing a $99.7\%$ reduction. Design~2 shows a similar reduction from $1.37K$ to $4.17$, while Design~3 drops from $1.43K$ to $4.39$, and Design~4 from $1.43K$ to $3.77$. In Design~6, the CFP improves from $1.55K$ to $4.86$, maintaining the same level of reduction.

Design~5 is the only outlier, where \emph{ECOLogic} incurs higher carbon cost—$3.24\text{K}$ versus $1.24\text{K}$—likely due to workload-specific dynamic power or a larger fabric footprint. \uline{Still, when averaged across Designs~1--4 and 6, \emph{ECOLogic} achieves a 99.68\% reduction in deployment carbon, corresponding to a $300$--$500\times$ improvement over FPGA-only solutions}~\cite{sudarshan2024eco,sudarshan2023greenfpga}. This efficiency is largely attributed to the reuse of the same silicon across multiple applications without requiring new tape-outs. Unlike full-FPGA platforms that suffer from significant area and power overhead, \emph{ECOLogic} limits reconfigurability to targeted IPs. This selective approach retains near-ASIC performance while minimizing operational emissions. As discussed in Sec.~\ref{sub:sus_analy} and Sec.~\ref{sub:cir_analy}, these reductions position \emph{ECOLogic} as a practical and environmentally sustainable choice for modern SoC deployment.

\subsection{Aging Analysis} \label{subsec:age}

\begin{figure}[t] % Positioning options: here, top, bottom, page
    \centering
    \includegraphics[width=0.45\textwidth]{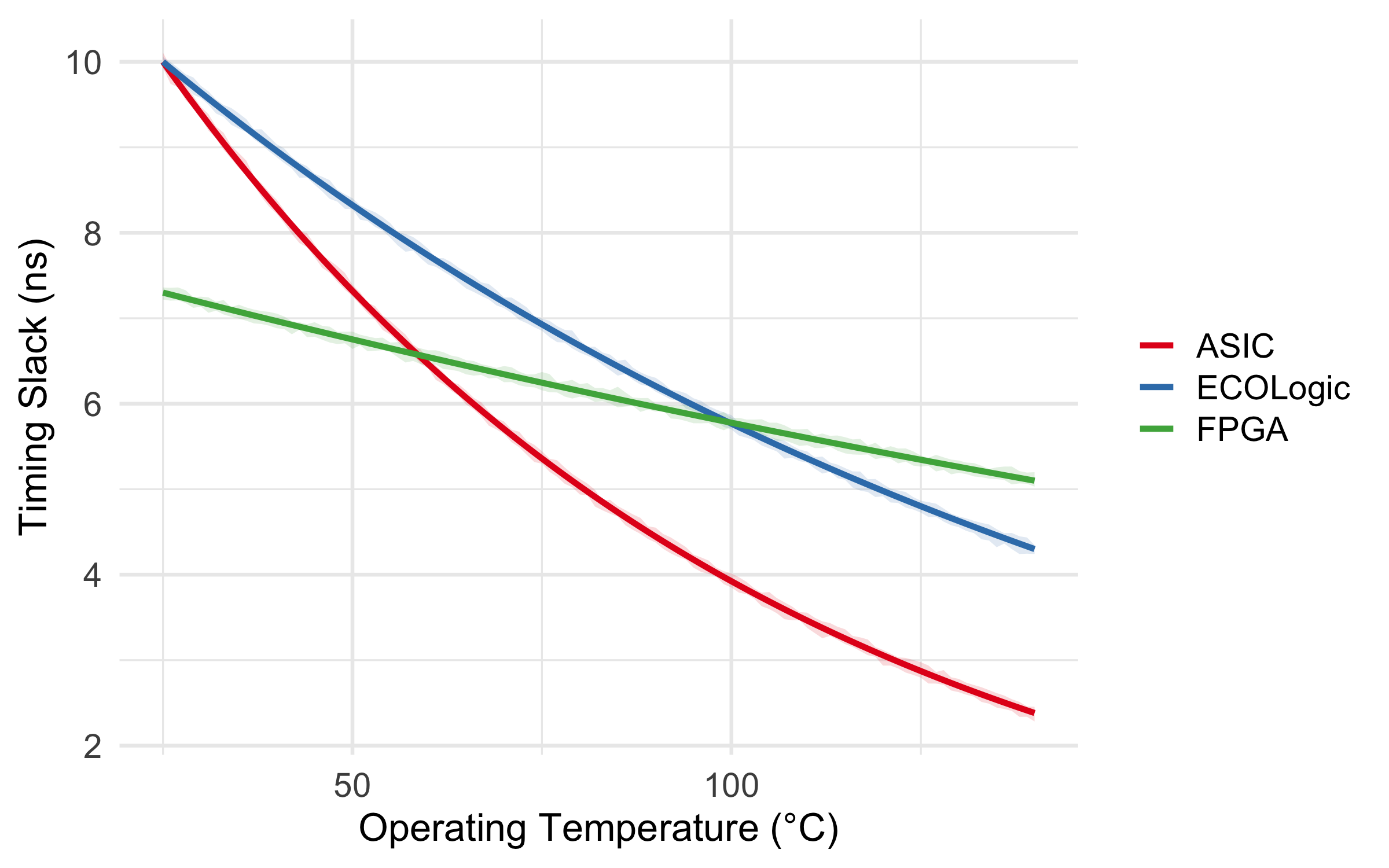}
    \caption{Timing slack degradation as a function of increasing temperature across ASIC, FPGA, and \emph{ECOLogic} platforms. While ASIC and FPGA exhibit sharp slack drops beyond 80\textdegree{}C, \emph{ECOLogic} maintains higher slack ($>5$\,ns at 130\textdegree{}C) due to its ability to dynamically remap logic to healthier fabric regions.}
    \label{fig:temp_slack} % Label for referencing the figure in your text
\end{figure}

\begin{figure}[b] % Positioning options: here, top, bottom, page
    \centering
    \includegraphics[width=0.5\textwidth]{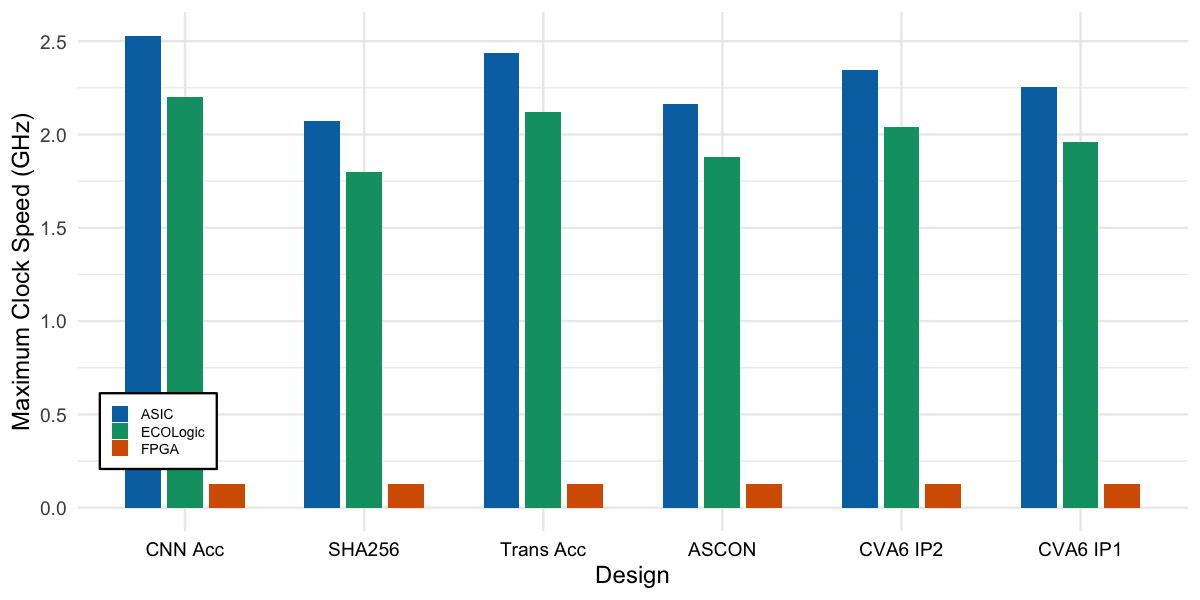}
    \caption{Comparison between maximum clock speed of different designs in ASIC, FPGA, and \emph{ECOLogic}.}
    \label{fig:perf_analysis} 
\end{figure}

\noindent Fig.~\ref{fig:temp_slack} shows the temperature-induced timing slack degradation across ASIC, FPGA, and \emph{ECOLogic}. Below $60^\circ$C, all platforms maintain stable slack ($>8$\,ns), indicating minimal aging impact. However, beyond $80^\circ$C, slack begins to deteriorate sharply—particularly in FPGA, which drops below $6$\,ns by $100^\circ$C. ASIC slack degrades more severely, reaching nearly $2$\,ns at $130^\circ$C, due to heightened vulnerability to aging effects such as Hot Carrier Injection (HCI) and Negative Bias Temperature Instability (NBTI)~\cite{kufluoglu2004computational}, which compromise drive strength and timing closure in critical paths. \uline{\emph{ECOLogic}, in contrast, demonstrates slower slack degradation—retaining $>5$\,ns at $130^\circ$C—by leveraging its ability to dynamically remap logic from thermally stressed, degraded regions to higher-performing, idle regions of the embedded eFPGA fabric.} This post-fabrication remapping restores timing margins without requiring chip re-spins. In thermally demanding domains like automotive and aerospace, this capability helps maintain reliable system operation over extended lifetimes. By adaptively reallocating aging logic to functional fabric resources, \emph{ECOLogic} prolongs chip usability and reduces the environmental and economic cost of premature replacement.

%\noindent Fig.~\ref{fig:temp_slack} highlights the impact of elevated operating temperatures on timing slack degradation. At nominal temperatures, slack remains relatively stable, indicating minimal influence from aging mechanisms on transistor-level parameters such as threshold voltage and drive strength. However, as temperature rises, a sharp and exponential decline in slack is observed—characteristic of aging phenomena like Hot Carrier Injection (HCI) and Negative Bias Temperature Instability (NBTI)~\cite{kufluoglu2004computational}. These effects, driven by interface state generation at the Si/SiO\textsubscript{2} boundary, gradually degrade transistor performance, ultimately eroding timing margins and threatening functional reliability under prolonged thermal stress. To mitigate such degradation, \emph{ECOLogic} leverages embedded reconfigurable fabric within the ASIC domain, enabling post-fab recovery of timing paths through dynamic logic remapping. As aging-induced shifts accumulate, critical logic blocks can be reassigned to idle, redundant fabric resources—restoring timing closure without requiring chip re-spins. This reconfiguration capability proves particularly beneficial in thermally harsh environments such as automotive or aerospace systems, where long-term reliability is paramount. By enabling adaptive resource reallocation in response to wear-out conditions, \emph{ECOLogic} offers a scalable mechanism to extend system lifespan while reducing the environmental and economic costs associated with premature hardware obsolescence.

\begin{figure}[t] % Positioning options: here, top, bottom, page
    \centering
    \includegraphics[width=0.5\textwidth,keepaspectratio]{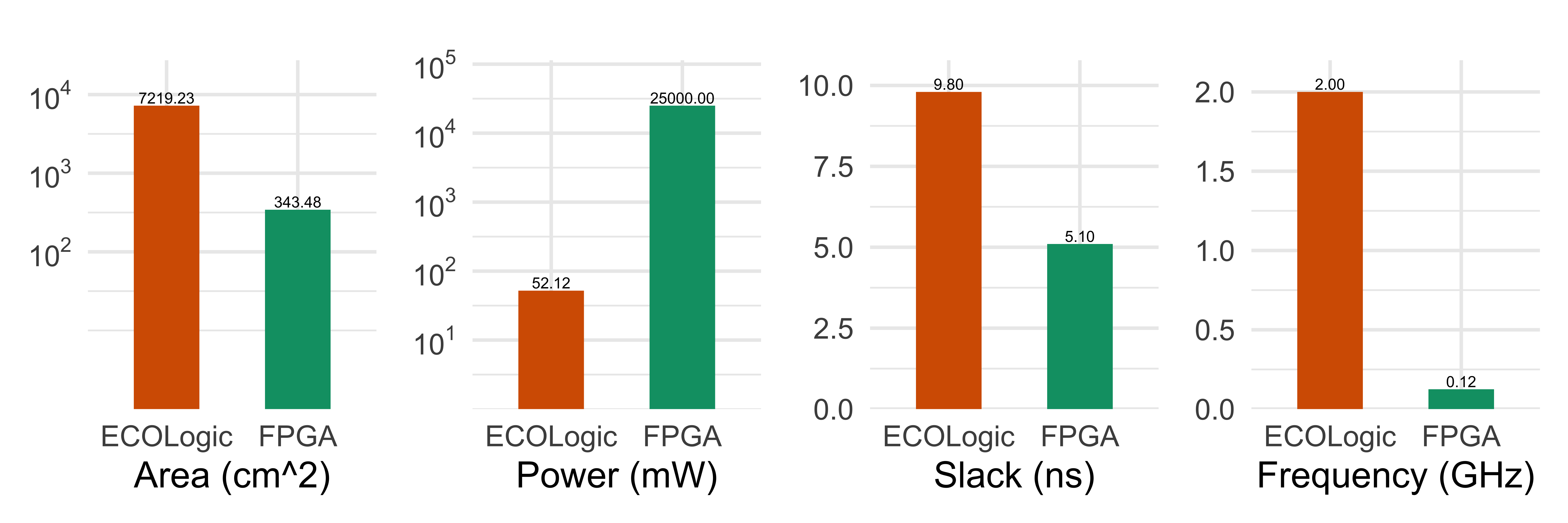}    
    \caption{Comparison of Area, Power, Timing Slack, and Frequency for \emph{ECOLogic} and FPGA.}
    \label{fig:comb_graph}
\end{figure}

\vspace{-0.1in}
\subsection{Performance Analysis}
\noindent Fig.~\ref{fig:perf_analysis} compares the maximum clock speeds across ASIC, FPGA, and \emph{ECOLogic} implementations for six IPs. FPGA designs consistently operate at just $0.125$\,GHz across all workloads, reflecting the inherent performance limitations of general-purpose reconfigurable logic. In contrast, ASIC implementations achieve between $2.07$\,GHz (SHA-256) and $2.53$\,GHz (CNN accelerator), with \emph{ECOLogic} closely tracking these values—ranging from $1.80$\,GHz to $2.20$\,GHz across the same designs. For instance, in the CNN accelerator, \uline{\emph{ECOLogic} sustains $2.20$\,GHz, just $13\%$ below ASIC and over $1600\%$ faster than FPGA.} This near-parity is made possible by hardening performance-critical logic in ASIC while reserving eFPGA regions for adaptable components. Thus, \emph{ECOLogic} preserves reconfigurability without compromising performance.

Fig.~\ref{fig:comb_graph} compares \emph{ECOLogic} and FPGA across area, power, slack, and frequency. \emph{ECOLogic} occupies a larger area ($7219$\,mm$^2$ vs. $343$\,mm$^2$) due to its embedded fabric but \uline{consumes approximately $480\times$ less power ($52$\,mW vs. $25{,}000$\,mW). Timing slack is also significantly higher: $9.8$\,ns for \emph{ECOLogic} versus $5.1$\,ns for FPGA. Frequency-wise, \emph{ECOLogic} achieves $2.0$\,GHz compared to $0.125$\,GHz in FPGA—a $16\times$ improvement.} These numbers confirm that \emph{ECOLogic} combines the adaptability of reconfigurable platforms with near-ASIC efficiency, making it a compelling choice for modern, performance-sensitive applications that still benefit from post-fab updates.

% \begin{comment}

% Following comparison between ASIC vs eFPGA vs FPGA:

% 1. PPA (Power, Performance, Area) (Gives an edge over pure FPGA)

% 2. CFP calculation. (gives an edge over ASIC)

% 3. Age Analysis. (gives an edge over ASIC)

% 4. (Optional) Balance between selection of IP based on results.

% \end{comment}

\section{Conclusion}\label{sec:conclusion}

% Summarize the contribution of the paper.
\noindent \emph{ECOLogic} presents a scalable, reconfigurable architecture that bridges the performance-efficiency gap between traditional ASICs and FPGAs, while advancing sustainability through reuse. By selectively embedding eFPGA fabric within ASIC systems, \emph{ECOLogic} enables post-fabrication updates, logic remapping under aging, and security through redaction. Across six diverse IPs, \emph{ECOLogic} achieves an average $99.7\%$ reduction in deployment carbon footprint compared to standalone FPGAs and improves clock frequency by $16\times$ while reducing power by over $480\times$. Timing slack improves by nearly $2\times$, enabling robust operation under thermal stress. These results demonstrate that \emph{ECOLogic} not only offers near-ASIC performance but also promotes long-term reliability and environmental efficiency—making it well-suited for future SoC designs in dynamic, multi-application environments.

\begingroup
\bibliographystyle{IEEEtran}
\bibliography{sample-base}
\endgroup

\end{document}